\pdfoutput=1
\documentclass[12pt]{article}
\usepackage{amsmath}

\usepackage{amssymb}
\usepackage{graphicx}
\usepackage{adjustbox}
\usepackage{multirow}
\usepackage{booktabs,caption}
\usepackage[flushleft]{threeparttable}
\usepackage{float}
\usepackage{color}
\newcommand{\be}{\begin{equation}}
\newcommand{\ee}{\end{equation}}
\newcommand{\bea}{\begin{eqnarray}}
\newcommand{\eea}{\end{eqnarray}}

\catcode`@=12

\begin{document}
\thispagestyle{empty}
\begin{center}
{\Large\bf
{Addressing the High-$f$ Problem in Pseudo - Nambu - Goldstone Boson Dark Energy Models with Dark Matter - Dark Energy Interaction }}\\
\vspace{1cm}
{{\bf Upala Mukhopadhyay}\footnote{email: upala.mukhopadhyay@saha.ac.in},
{\bf Avik Paul}\footnote{email: avik.paul@saha.ac.in},
{\bf Debasish Majumdar}\footnote{email: debasish.majumdar@saha.ac.in}}\\
\vspace{0.25cm}
{\normalsize \it Astroparticle Physics and Cosmology Division,}\\
{\normalsize \it Saha Institute of Nuclear Physics, HBNI,} \\
{\normalsize \it 1/AF Bidhannagar, Kolkata 700064, India}\\
\vspace{1cm}
%%%%%%%%%%%%%%%%%%%%%%%%%%%%%%%%%%%%%%%%%%%%%%%%%%
%{\bf ABSTRACT}
%%%%%%%%%%%%%%%%%%%%%%%%%%%%%%%%%%%%%%%%%%%%%%%%%%
\end{center}
\begin{abstract}
%20190720:
We consider a dark energy scenario driven by a scalar field $\phi$ with a pseudo - Nambu - Goldstone boson (pNGB) type potential $V(\phi)=\mu^4 \left( 1+ {\rm cos}(\phi/f) \right)$. The pNGB originates out of breaking of spontaneous symmetry at a scale $f$ close to Planck mass $M_{\rm{pl}}$. We consider two cases namely the quintessence dark energy model with pNGB potential and the other, where the standard pNGB action is modified by the terms related to Slotheon cosmology. We demonstrate that for this pNGB potential, high-$f$ problem is better addressed when the interaction between dark matter and dark energy is taken into account and that Slotheon dark energy scenario works even better over quintessence in this respect. To this end, a mass limit for dark matter is also estimated.  
\end{abstract}
%\hspace{1cm}\keywords{Dark Energy, Axion}
\newpage
\section{Introduction}
The observational data \cite{observation1} - \cite{observation2} reveal that our Universe is not only expanding with time but is undergoing an accelerated expansion. One of the most challenging problems of modern cosmology is to explain such a late time cosmic acceleration. The general conjecture is that a mysterious component of the Universe with negative pressure, broadly known as dark energy (DE) is responsible for the recent acceleration that accounts about 69$\text{\%}$  of the total mass energy content of the present Universe. In theory, the commonly used candidate for dark energy is the cosmological constant $\Lambda$  \cite{lambda CDM, kazuharu_review} in the Einstein's equations introduced by Einstein. The popular and widely used dark energy model namely $\Lambda$CDM model that includes cosmological constant $\Lambda$ and the cold dark matter (CDM) has some unsolved theoretical problems like the fine-tuning problem \cite{lambda CDM} and cosmic coincidence problem \cite{cosmic coincidence}. 

Another concept to address the dark energy is to assume the existence of a scalar field $\phi$ with a slow rolling potential $V(\phi$) as the source of dark energy. This scalar field $\phi$, named as quintessence field, provides the dynamical nature of the dark energy, in contrast to the cosmological constant explanation according to which the dark energy is constant throughout the evolution of the Universe. Extensive studies have been done to study the nature of quintessence dark energy model \cite{quint1,{quint2}}. A well motivated alternative description of dark energy is given by modified gravity models of dark energy, inspired by the theories of extra dimensions \cite{extra dim1,{extra dim2}}. It is observed that to obey the observational results, the scalar field $\phi$ should possess a very flat potential $V(\phi)$ and very light mass.

The other dark sector component of the Universe is dark matter (DM) which contains about $26.5\text{\%}$ of the total mass - energy content of the present Universe. The existence of the dark matter is confirmed by various observational results \cite{dm1} - \cite{dm2}. Various attempts have been made in the literature to explain the unknown dark matter in theories beyond Standard Model of particle physics. But in this work where we addressed dark matter - dark energy interactions, the dark matter is considered to be a scalar field.

Till date there are no theoretical considerations or experimental observations that seem to suggest that a possibile dark matter - dark energy interaction can not exist. We consider here a nonminimal coupling between the two dark sector components namely, dark matter and dark energy instead of treating them independently. In literature various authors discussed the interacting dark energy (IDE) models \cite{Wang:2016} to address different phenomenological problems \cite{Farrar:2004} - \cite{DiValentino:2019ffd}. 

As mentioned, a popular dark energy model is scalar field dark energy model where one considers a slowly varying potential (slow roll) for the scalar field which the latter tracks as the potential changes over time. Even though the idea of the scalar field dark energy model is well motivated but in Ref. \cite{267_sami} Kolda and Lyth pointed out a serious problem for any slow rolling scalar field dark energy model. The problem is to prevent any additional terms to the field potential $V(\phi)$, which would spoil the flatness of the potential. In order to avoid this problem, the dark energy models are considered where the light mass of the scalar field $\phi$ is protected by a symmetry. Such scenario can be realised if a pseudo - Nambu - Goldston boson (pNGB) acts as a dynamical dark energy field. This concept is studied in Refs. \cite{279_sami} and \cite{280_sami}.   

The spontaneous symmetry breaking of a global $U(1)$ symmetry results in producing a massless Goldstone boson. Such spontaneous symmetry breaking leads to two modes, one is the massive radial mode  and the other is the massless angular mode (named as NGB) at the symmetry breaking energy scale. A pseudo - Nambu - Goldstone boson is produced when the NGB acquires mass at the soft explicit symmetry breaking scale which is lower than the spontaneous symmetry breaking scale. A popular example of pNGB is Axion \cite{Peccei:2006as} - \cite{khlopov}. As mentioned, pNGB could play the role of the quintessence dark energy field with the following form of the potential \cite{279_sami}
\begin{equation}
V(\phi)=\mu^4 \left( 1+ {\rm cos}(\phi/f) \right)\,\,.\label{intro}
\end{equation}
In the above, $f$ is the spontaneous global symmetry breaking scale which controls the steepness of the potential and $\mu$ is the explicit global symmetry breaking scale. The above potential is periodic with period $2\pi f$ and the field value $\phi$ ranges from $0$ to $2\pi f$. This periodic potential is special because it stabilises the mass from quantum corrections \cite{267_sami} and suppresses the fifth-force constraints \cite{carrol}. In Refs. \cite{Frieman} - \cite{ArkaniHamed:2003mz} the authors found that  generally for the quintessence dark energy with pNGB potential, $f\geqslant M_{\rm pl}$, $M_{\rm pl}$ being the reduced Planck mass and indicated that the larger value of $f$ corresponds to the flatter potential. But it is very difficult to interpret such high value of $f$ since such values of $f$ are not compatible with the valid domain of field theory. For this range, quantum gravity corrections can not be controlled \cite{ArkaniHamed:2003mz}. This problem is termed as high-$f$ problem of pNGB quintessence dark energy model. Few authors have attempted to solve this high-$f$ issue earlier \cite{Kaloper:2005aj, Adak:2014moa}. For example in Ref.  \cite{Kaloper:2005aj} this has been suggested that $N$ number of pNGB fields drive the late time acceleration and in Ref. \cite{Adak:2014moa} the authors discussed the issue by adding extra phenomenological terms to the quintessence lagrangian. Here in this work we explore a new approach where we consider interacting dark energy (IDE) model and  
the nonminimally coupled dark matter-dark energy scenario to address the high-$f$ value problem  of pNGB quintessence.

Dark matter - dark energy (DM-DE) interactions are sometimes marked by considering both dark energy and dark matter to be fluids or both of them to be scalar fields or even one of them is a scalar field and other is fluid. In this work we take both the dark energy and dark matter as real scalar fields and treat the interactions in a way which is more fundamental in nature. Here, we consider two dark energy models, the standard quintessence dark energy model and the Slotheon field dark energy model \cite{me_sloth} - \cite{178_thesis}. The Slotheon dark energy model is a modified gravity model, inspired by extra - dimensional theories. This follows from Dvali Gabadadze Porrati (DGP) model \cite{DGP} (an extra dimensional model) which in its decoupling limit can be described by a scalar field named as Galileon field in Minkowski spacetime and the field obeys Galileon shift symmetry \cite{173_thesis, 176_thesis}. The generalisation of the Galileon shift symmetry to curved spacetime leads to the Slotheon field \cite{178_thesis}. In literature there are references where it is shown that in some cases such dark energy models fit better  with the observational or theoretical constraints than the standard quintessence model \cite{me_sloth, me_swamp, gal_swamp, gal_pert}. Therefore in this work too we evaluate our results for each of these two cases (namely quintessence dark energy and Slotheon field dark energy) and compare them. For both the cases we consider the nonminimal coupling between dark matter and dark energy and furnish our results for both the scenarios with and without DM-DE interactions.

This paper is organised as follows. In section 2 we briefly describe in general the dark matter-dark energy interactions. Section 3 is devoted to explain the coupled (coupled to dark matter) quintessence dark energy model  with pNGB potential while in section 4, coupled Slotheon field dark energy model with pNGB potential is described. We present our calculations and results in section 5 and finally in section 6 a summary and discussions are given.
 
\section{Dark Energy - Dark Matter Interaction}
In this section we briefly furnish the formalism for the DM-DE interactions that is followed in this work.

In order to describe the DM-DE interactions the energy-momentum conservation equations are written with a introduction of an exchange term $Q$ as,
\begin{eqnarray}
\dot{\rho}_{\rm DE} + 3 H (\rho_{\rm DE} + p_{\rm DE})&=&Q\,\,,  \label{interaction_de}\\
\dot{\rho}_{\rm DM} +3 H \rho_{\rm DM} &=& -Q \,\,,\label{inetraction_dm}
\end{eqnarray}
where $\rho_{\rm DE}$ and $\rho_{\rm DM}$ are the energy densities of dark energy and dark matter respectively, $H$ denotes the Hubble parameter while $\dot{A}$ represents time derivative of the variable $A$. In the above, $Q$ refers to the energy transfer between dark matter and dark energy and defines the DM-DE interaction coupling. In literature there are several studies of this interaction where  the dark sector is considered to be a two component fluid with some usual forms for coupling $Q$ are adopted \cite{usual_Q1, usual_Q2}. But in interacting quintessence model where dark energy is assumed to be a scalar field $\phi$, the coupling $Q$ is given as \cite{interacting_quint1, interacting_quint2},
\begin{equation}
Q=f(\phi) \dot{\phi} \rho_ b \,\,,\label{Q}
\end{equation}
where $f(\phi)$ is a function of field $\phi$ and $\rho_b$ is the energy density of the background fluid. Substituting $\rho_b$ by $\rho_{\rm DM}$ in eq. (\ref{Q}) one may obtain the DM-DE interaction coupling $Q$ in a model where dark matter interacts with the quintessence dark energy field \cite{44_lisboa}. In Ref. \cite{lambda CDM} the authors have explicitely derived the form of $Q$ in eq. (\ref{Q}). They have considered a general 4-D action of a scalar field $\phi$ interacting with matter fluid (of energy density $\rho_b$) and obtained the form of $Q$ as $Q=f(\phi) \dot{\phi} \rho_ b$ where the expression of $f(\phi)$ depends on the nature of the coupling between $\phi$ and matter. Moreover, they have also shown that by rewriting (carrying out a conformal transformation on the metric $g_{\mu \nu}$) the lagrangian %of a scalar field $\phi$ and a barotropic fluid 
from Jordan frame to Einstein frame, one can obtain the same form of coupling $Q$ (eq. (\ref{Q})).

A more fundamental approach to describe the DM-DE interactions is to treat both the dark matter and dark energy as fields \cite{Farrar:2003uw, Micheletti:2009pk, PRD_formalism}. In the present work both the fields are chosen as real scalar fields similar to what is considred in Ref. \cite{PRD_formalism}.
The action contains a kinetic term for dark energy field $\phi$, a kinetic term for dark matter field $\chi$ and a total potential $V(\phi,\chi)$, where $V(\phi,\chi)$ includes the potentials for dark energy and dark matter. The action for such a system can be written as,  
\begin{equation}
S=\int d^4x \sqrt{-g} \left[ -\frac{1}{2} g^{\mu\nu} \left( \partial_\mu \phi \partial_\nu \phi +\partial_\mu \chi \partial_\nu \chi \right) - V(\phi,\chi) \right]\,\,.\label{action}
\end{equation}
In the above, $g^{\mu\nu}$ denotes the metric tensor and $g$ is the determinant of the metric tensor. These models also lead to the coupling of interacting quintessence case (eq. (\ref{Q})) when the interaction is of the form of dark matter mass term. %This approach being more elementary we follow in this work this formalism for DM-DE interaction when both the dark matter and dark energy are considered to be real scalar fields. 
The formalism for this DM-DE interaction model is studied in Ref. \cite{PRD_formalism} with $V(\phi, \chi)=V_{\rm DE}(\phi) + V_{\rm DM}(\phi, \chi)$ where $V_{\rm DM}(\phi, \chi)=\frac{1}{2} M^2(\phi) \chi^2$ and $M^2(\phi)$ is expressed in terms of ``bare mass" $m$ of the dark matter and $F(\phi)$ signifies DM-DE interaction. The mass function $M^2(\phi)$ is given as $M^2(\phi)=m^2+F(\phi)$. With $\rho_{\rm DM}$ derived as $\rho_{\rm DM}(\phi,a)=n_o a^{-3} M(\phi)$ and $V_{\rm eff}$ defined as $V_{\rm eff}(\phi,a)=V_{\rm DE}(\phi)+\rho_{\rm DM}(\phi,a)$, we consider in this work two scenarios namely coupled quintessence dark energy and coupled Slotheon dark energy.

\section{Coupled Quintessence Dark Energy Model}
The action for coupled quintessence scenario where the standard quintessence scalar $\phi$ is the dark energy canditate is written as
\begin{equation}
S=\int d^4x \sqrt{-g} \left(\frac{1}{2} M_{\rm pl}^2 R -\frac{1}{2} g^{\mu\nu} \left(\partial_\mu \phi \partial_\nu \phi + \partial_\mu \chi \partial_\nu \chi    \right)-V_{\rm DE}(\phi) -\frac{1}{2} m^2 \chi^2-\beta \phi^2\chi^2\right) + S_m + S_r\,\,, \label{action_quint}
\end{equation}
where $M_{\rm pl}$ and $R$ are reduced Planck mass and Ricci scalar respectively and $\beta$ denotes the coupling constant for DM-DE interaction. In the above $S_m$ and $S_r$ refer to the action of baryonic matter and action of radiation component respectively. We consider the quintessence field with pNGB potential for our dark energy model, therefore the expression for the potential $V_{\rm DE}(\phi)$ is similar as in eq. (\ref{intro})
\begin{equation}
V_{\rm DE}(\phi)=\mu^4 \left( 1+ {\rm cos}(\phi/f) \right)\,\,.\label{pngb}
\end{equation}
Hence in our case the effective potential is given by
\begin{equation}
V_{\rm eff}(\phi,a)=\mu^4\left(( 1+ {\rm cos}(\phi/f) \right) + n_o a^{-3}M(\phi)\,\,,\label{veff_quint}
\end{equation}
where symbols have similar meanings as described in the previous section and here $M^2(\phi)$ is given by
\begin{equation}
M^2(\phi)=m^2+ 2 \beta \phi^2\,\,. \label{M}
\end{equation}
By varying the action of eq. (\ref{action_quint}) with respect to the metric, the Friedmann equations are obtained as
%\begin{eqnarray}
$3 M_{{\rm pl}}^2 H^2 = \rho_m$ $+\rho_r + \dfrac{\dot{\phi}^2}{2}$ $+ V_{\rm eff}(\phi,a)$ ($\rho_m$ and $\rho_r$ are mass and radiation densities respectively),
$M_{{\rm pl}}^2 (2\dot{H} + 3H^2) = -\frac{\rho_r}{3}$ $-\dfrac{\dot{\phi}^2}{2} +$ $V_{\rm eff}(\phi,a)$ ($H$ is the Hubble parameter) and
$0 = \ddot{\phi} $ $+ 3H\dot{\phi} + \frac{\partial V_{\rm eff}(\phi,a)}{\partial \phi}$\,. %\label{EE3 quint}
%\end{eqnarray}

These are solved by constructing the autonomous set of equations in terms of a chosen set of dimensionless variables 
$x = \dfrac{\dot{\phi}}{\sqrt{6}H M_{{\rm pl}}}$,
$y = \dfrac{\sqrt{V_{\rm DE}(\phi)}}{\sqrt{3} H M_{{\rm pl}}}$,
$\lambda = -M_{{\rm pl}} \dfrac{\frac{dV_{\rm DE}(\phi)}{d\phi}}{V_{\rm DE}(\phi)}$ and
$\xi =  \frac{\phi}{m}$
as 
$$
\dfrac{dx}{dN} = \frac{P}{\sqrt{6}}-x \frac{\dot{H}}{H^2},\,\,\,\,\, 
\dfrac{dy}{dN} = -y \left(\sqrt{\dfrac{3}{2}} \lambda x + \dfrac{\dot{H}}{H^2}\right)\,,
$$
\begin{equation}
\dfrac{d\lambda}{dN} = -\sqrt{6} x \lambda^2 \left(\frac{V_{\rm DE} \frac{d^2 V_{\rm DE}}{d \phi^2}}{\left(\frac{dV_{\rm DE}}{d\phi}\right)^2}-1 \right),
\,\,\,\, \dfrac{d\xi}{dN} = \sqrt{6} \frac{x}{b}\,,\label{auto4 quint} 
\end{equation}
with $N={\rm ln}$ $a$ is the number of e-foldings, $b=\frac{m}{M_{\rm pl}}$ and
$\frac{\dot{H}}{H^2} = \frac{1}{2} \left(\frac{3 \Omega_{\rm do}}{a^3}-3 x^2+3 y^2-\Omega_r-3\right)$\,,
$P = -\frac{6 \beta  \xi  \Omega_{\rm do}}{a^3 \left(2 b \beta  \xi ^2+b\right)}-3 \sqrt{6} x+3 \lambda  y^2$\,\,\,,
$\frac{V_{\rm DE} \frac{d^2 V_{\rm DE}}{d \phi^2}}{\left(\frac{dV_{\rm DE}}{d\phi}\right)^2} = \frac{1}{2}\left(1-\frac{1}{\lambda^2 (\frac{f}{M_{\rm pl}})^2}\right),$ 
where the other notations carry usual significance and $\Omega_{\rm do}$ is the dark matter density parameter at the present epoch. 

The effective equations of state parameter $\omega_{\rm eff}$ and equation of state parameter of dark energy
$\omega_{\rm DE}$ are now obtained as 
\begin{equation}
\omega_{\rm eff} = -1-\frac{2 \dot{H}}{3 H^2}\,,\,\,\,\,\, 
\omega_{\rm DE} = \frac{\omega_{\rm eff}-\frac{\Omega_r}{3}}{\Omega_{\rm DE}}\,. \label{omega}
\end{equation}

%It is needless to mention here that density parameters $\Omega_x$ and equation of state parameter $\omega_{\rm DE}$ are now obtained in terms of the dimensionless variables of eqs. (\ref{x quint} - \ref{xi quint}) and the evolutions of these cosmological parameters can easily be derived by solving the set of coupled equations (eqs. (\ref{auto1 quint} - \ref{auto4 quint})) with proper initial conditions.

\section{Coupled Slotheon Dark Energy Model}
The Slotheon field model \cite{178_thesis} is a scalar field model which is classified as a modified gravity model of dark energy. The Slotheon field model is inspired by Dvali Gavadadze Poratti (DGP) model \cite{DGP} - an extra dimensional model with one extra dimension. The DGP model in its decoupling limit $r_c \rightarrow \infty$ \cite{24_swamp, 25_swamp} ($r_c$ seperates 4-D and 5-D regimes and defined as $r_c=\frac{M_{\rm pl}^2}{2 M_5^2}$, where $M_{\rm pl}$ and $M_5$ are bulk and brane Planck masses respectively) can be described by a scalar field (say $\pi$), dubbed as Galileon field \cite{173_thesis, 176_thesis}. Here the field $\pi$ respects a shift symmetry known as Galileon shift and is given by $\pi \longrightarrow \pi + a + b_\mu x^\mu$ \cite{177_thesis}. The symbols $a$ and $b_\mu$ represent a constant and a constant vector respectively. The Slotheon field arises when this Galileon transformation is generalised to curved spacetime \cite{178_thesis} and the Slotheon field obeys this curved Galileon transformation, 
\begin{equation}
\pi(x) \rightarrow \pi(x)+c+c_a\int_{\Gamma,x_0}^x \gamma^a\,\,,
\end{equation}
where $\gamma^a$ is a set of Killing vectors, $x_0$ is a reference point connected to another point $x$ through the curve $\Gamma$ while $c$ and $c_a$ are respectively a constant and a constant vector. It is obsereved in Ref. \cite{me_sloth} and \cite{deba_sloth} that if Slotheon field model of dark energy is considered, then the slow rolling criteria will be more favoured than the standard quintessence dark energy model. This is because the former induces an extra friction which favours the slow rolling nature of the field. Moreover in Ref. \cite{me_swamp} it is demonstrated that the Swampland criteria are better satisfied with the Slotheon field model of dark energy over the quintessence model.

In this section we consider the Slotheon field $\pi$ as dark energy and explore the behaviours of different cosmological parameters when it is coupled to the dark matter. The action of the coupled Slotheon dark energy field is given as
%\begin{equation}
\begin{eqnarray}
%\end{eqnarray}
S=\int d^4x \sqrt{-g} \left(\frac{1}{2} M_{\rm pl}^2 R -\frac{1}{2} g^{\mu\nu}\partial_\mu \pi \partial_\nu \pi + \frac{G^{\mu\nu}}{2 M^2}\partial_\mu \pi \partial_\nu \pi -\frac{1}{2} g^{\mu\nu} \partial_\mu \chi \partial_\nu \chi -V_{\rm DE}(\pi) -\frac{1}{2} m^2 \chi^2-\beta \pi^2\chi^2\right) \nonumber \\ + S_m + S_r\,\,.  \hspace{15cm} \label{action_sloth}
\end{eqnarray}
%\end{equation}
In the above $G^{\mu\nu}$ is the Einstein tensor and $M$ represents an energy scale and all the symbols are same as in eq. (\ref{action_quint}). It can be noted here that without the term $\frac{G^{\mu\nu}}{M^2}\partial_\mu \pi \partial_\nu \pi$ in the action of eq. (\ref{action_sloth}), both the actions of eqs. (\ref{action_quint}) and (\ref{action_sloth}) are identical. In the Slotheon case also we have similar form of dark energy potentials as in eqs. (\ref{pngb} -  \ref{M}) by replacing $\phi$ by the Slotheon field $\pi$
(e.g. $V_{\rm DE}(\pi)=\mu^4 \left( 1+ {\rm cos}(\pi/f) \right)$ and similarly for $V_{\rm eff}(\pi,a)$ and $M^2(\pi)$).

The Friedmann equations for the action in  eq. (\ref{action_sloth}) can now be derived as
\begin{eqnarray}
3 M_{{\rm pl}}^2 H^2 &=& \rho_m +\rho_r + \dfrac{\dot{\pi}^2}{2} + \dfrac{9 H^2 \dot{\pi}^2}{2M^2} + V_{\rm eff}(\pi, a)\,\,, \label{EE1 sloth}\\
M_{{\rm pl}}^2 (2\dot{H} + 3H^2) &=& -\frac{\rho_r}{3}-\dfrac{\dot{\pi}^2}{2} +V_{\rm eff}(\pi,a) + (2\dot{H} + 3H^2) \dfrac{ \dot{\pi}^2}{2M^2} + \dfrac{2 H\dot{\pi}\ddot{\pi}}{M^2}\,\,, \label{EE2 sloth}\\
0 &= & \ddot{\pi} + 3H\dot{\pi} + \dfrac{3 H^2}{M^2}\left(\ddot{\pi} + 3H\dot{\pi} + \dfrac{2\dot{H}\dot{\pi}}{H}\right) + \frac{\partial V_{\rm eff}(\pi,a)}{\partial \pi}\,\,. \label{EE3 sloth}
\end{eqnarray}
The Friedmann equations above for Slothean field $\pi$ are solved by constructing a set of differential equations in terms of a suitable set of dimensionless variables as described for quintessence field 
in Sect. 3. But here, we need to define one more dimensionless variable $\epsilon = H^2/2M^2$ and an additional differential equation 
$\frac {d\epsilon} {dN} = 2\epsilon \frac {\dot{H}}{H^2}$. This additional variable $\epsilon$ arises due to the term  $\frac{G^{\mu\nu}}{M^2}\partial_\mu \pi \partial_\nu \pi$ in the action of eq. (\ref{action_sloth}). For the Slothean field the quantities $P$ and $\frac {\dot{H}}{H}$ (in the set of autonomous equations (see Sect. 3)) are 

\begin{eqnarray}
\frac{\dot{H}}{H^2} &=& \frac{-x^2 (6 \epsilon +1) (18 \epsilon +1)+4 \sqrt{6} x \epsilon  \left(\lambda  y^2-\frac{2 \beta  \xi \Omega_{\rm do}}{a^3 \left(2 b \beta  \xi ^2+b\right)}\right)}{4 \epsilon  \left(x^2 (1-18 \epsilon )-1\right)-\frac{2}{3}}\\
&& + \frac{\frac{(6 \epsilon +1) \left(a^3 \left(3 y^2-\Omega_r-3\right)+3 \Omega_{\rm do}\right)}{3 a^3}}{4 \epsilon  \left(x^2 (1-18 \epsilon )-1\right)-\frac{2}{3}} \nonumber \,\,,\\
P &=& \frac{6 \Omega_{\rm do} \left(\beta  \xi  \left(6 x^2 \epsilon -1\right)-3 \sqrt{6} b x \epsilon  \left(2 \beta  \xi ^2+1\right)\right)}{a^3 b \left(2 \beta  \xi ^2+1\right) \left(6 \epsilon  \left(x^2 (18 \epsilon -1)+1\right)+1\right)}\\
&& + \frac{3 a^3 b \left(2 \beta  \xi ^2+1\right)12 \sqrt{6} x^3 \epsilon -6 \lambda  x^2 y^2 \epsilon +\sqrt{6} x \left(-6 y^2 \epsilon +2 \Omega_r \epsilon -1\right)+\lambda  y^2}{a^3 b \left(2 \beta  \xi ^2+1\right) \left(6 \epsilon  \left(x^2 (18 \epsilon -1)+1\right)+1\right)}\,\,, \nonumber \label{p_sloth}
\end{eqnarray}
where all the symbols have their meaning as mentioned earlier. The effective equation of state parameter $\omega_{\rm eff}$  and the equation of state of dark energy $\omega_{\rm DE}$ for Slotheon field dark energy model can now be constructed using eqs. (\ref{EE1 sloth}) - (\ref{EE3 sloth}) and they will be of the same forms as in eq. (\ref{omega}). The variations of cosmological parameters $\Omega_x$ and $\omega_{\rm DE}$ are obtained from solving these equations with properly chosen initial conditions.

%We obtain from section 3, section 4 and section 5 that $\omega_{\rm DE}$ for coupled quintessence scenario and $\omega_{\rm DE}$ for coupled Slotheon dark energy scenario are time dependent variables. Therefore these two dynamical dark energy models would show different statefinder trajectories than $r=1$ and $s=0$. %Moreover, due to the different nature of the two models the pair ${r,s}$ would provide different trajectories for these two models and hence these two dynamical dark energy models can further be distinguished with the help of pair ${r,s}$. 
%Moreover, as evolutions of $\Omega_{\rm DE}$ and $\omega_{\rm DE}$ depend on dark energy - dark matter coupling, the non-minimally coupled nature of the quintessence model and Slotheon model would also be reflected in these trajectories. Variations of statefinder trajectories for the presence of dark energy - dark matter interaction are discussed in Ref. \cite{sfdmde}.}

\section{Calculations and Results}
In this section we furnish the results we obtain from solving the equations in section 3 and section 4.

\begin{figure}[H]
\begin{center}
\includegraphics[scale=0.8]{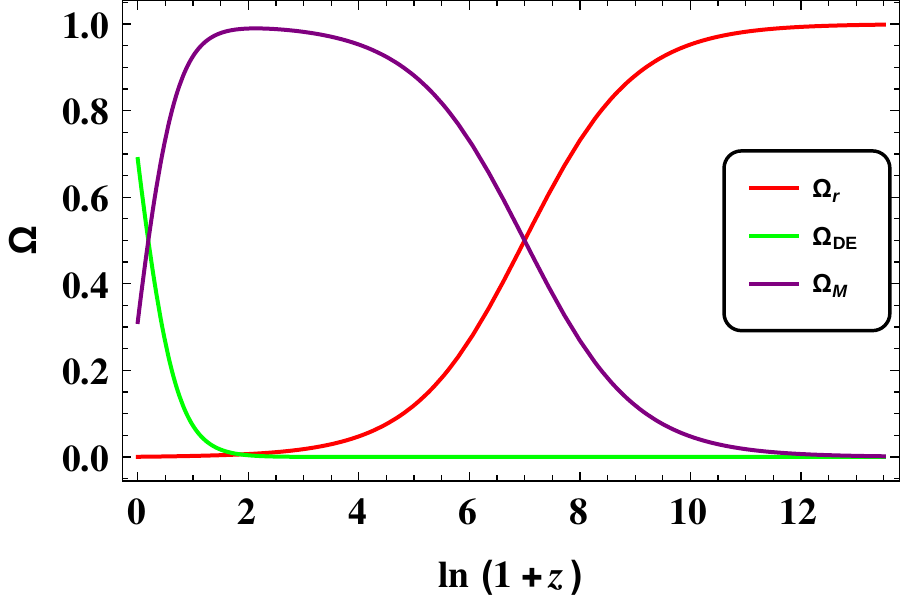} 
\caption{Variations of density parameters with redshift}\label{density}
\end{center}
\end{figure}

In Fig. \ref{density} the evolutions of the density parameters ($\Omega$) of different components of the Universe as functions of redshift $z$ are shown. In this figure dynamics of the density parameters for radiation, matter and dark energy are plotted by considering the quintessence dark energy model coupled to the dark matter field. From Fig. \ref{density} it can be observed that at a much later stage when ${\rm ln}(1+z) \simeq 0.4$, or $z \simeq 0.49$ the dark energy 
begins to dominate and the Universe enters into the phase of late time acceleration.
%It can also be noted from the figure that for the coupled dark energy model considered here, i.e., the dark energy is coupled only to dark matter (and no coupling with baryonic matter), we have a Universe which at early stage has a radiation era then followed by  matter dominated era and more recently experiencing the onset of an era of dark energy domination. These types of scenarios are well studied in Ref. \cite{Amendola:1999er}-\cite{baryon_lucha}.

\begin{figure}
\begin{center}
\includegraphics[scale=0.8]{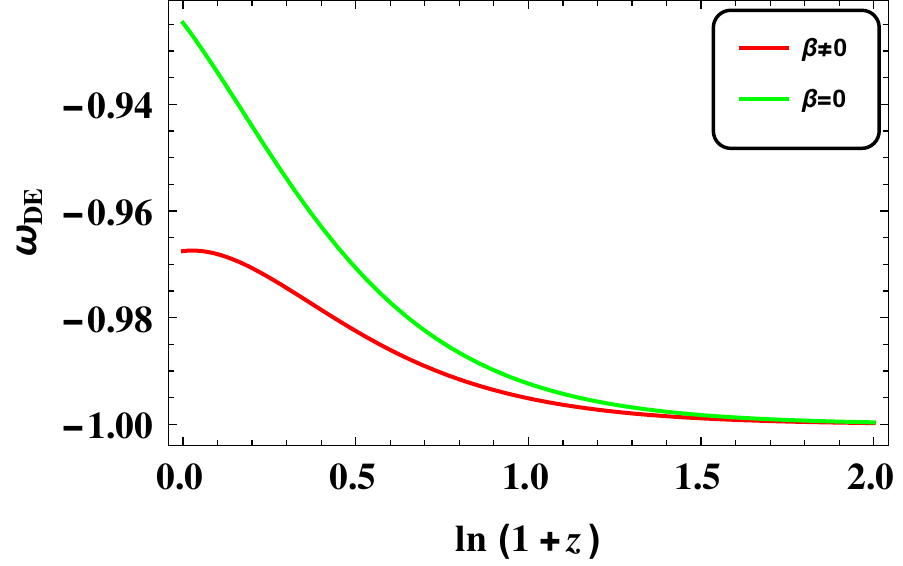}
\caption{Evolutions of dark energy equation of state parameter with redshift for quintessence dark energy model with DM-DE interaction (red line) and in the absence of DM-DE interaction (green line).}\label{eos1}
\end{center}
\end{figure}

In Fig. \ref{eos1} we show the variations of the dark energy equation of 
state parameter $\omega_{\rm DE}$  with redshift $z$ for standard 
quintessence dark energy model with potential given in eq. (\ref{pngb}). 
Fig. \ref{eos1} also indicates the thawing nature \cite{thawing} of the 
dark energy, i.e., $\omega_{\rm DE}$ is equal to $-1$ at early epochs 
which gradually deviates from the $-1$ with time, is clear. In 
Fig. \ref{eos1} we compare the variations for $\omega_{\rm DE}$ with $z$ 
when the DM-DE interactions are considered (red line in the 
plot, $\beta \neq 0$) with the same with no such interactions 
are included (the green line in the plot, $\beta = 0$). It is interesting 
to note that quintessence dark energy is more akin to $\Lambda$CDM 
model (where $\omega_{\rm DE} =-1$, 
throughout the time of evolution) when the dark energy field is coupled 
to the dark matter field. In this case and in the following cases 
we take $\beta=0.01$, $f=M_{\rm pl}$, $m=1$ keV if not otherwise stated. 
We may also add here that in this work we do not propose or 
adopt any particular 
particle dark matter theory or theories for the choice of a viable dark 
matter candidate. As the purpose of this work is to explore whether the
introduction of a dark energy-dark matter interaction term would be 
helpful in addressing the high $f$ problem for the pNGB potential 
of periodic nature (axion type $\sim (1+\cos(\phi/f))$),
we simply adopt here a dark matter with mass $m$ 
assuming this to be viable candidate. 
%It may be mentioned here that our results are independent of the 
%value of $\beta$ and only affected by the presence or absence of 
%the DM-DE interactions in the framework. This can be understood from 
%equations of section 3 and section 4. One observes from these equations 
%that the parameter $\beta$ cancels out from the numerator and the 
%denominator for the choices of initial conditions and values of other 
%parameters, appropriate for this work. 

\begin{figure}[H]
\begin{center}
\includegraphics[scale=0.8]{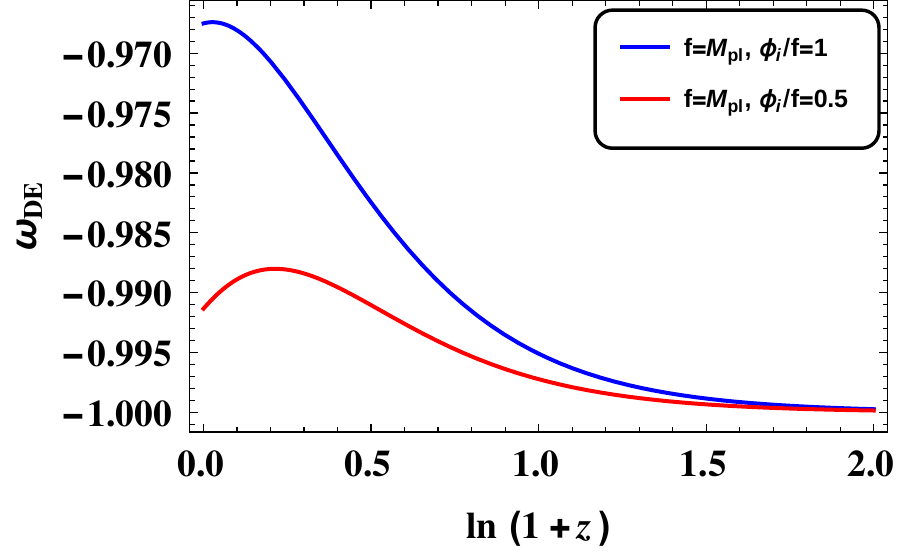}
\includegraphics[scale=0.8]{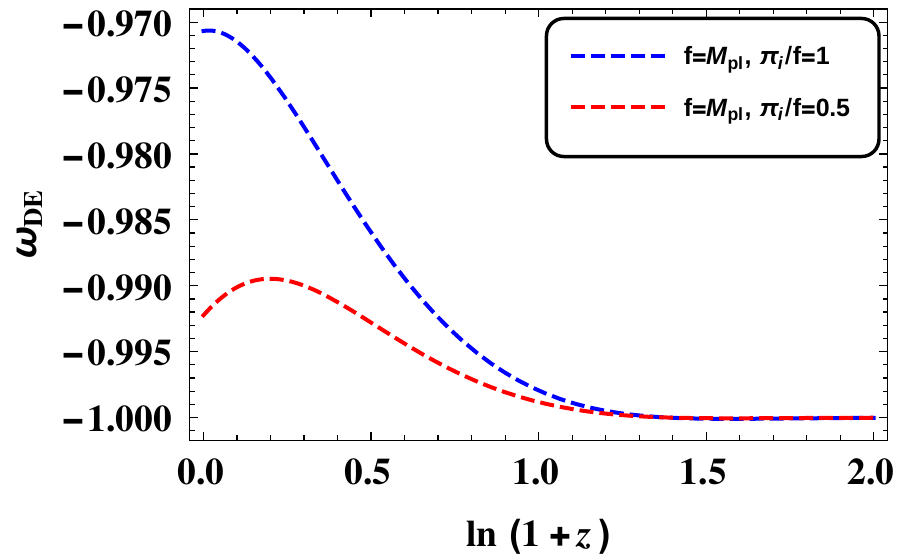}
%\includegraphics[scale=0.5]{ratiopi3.png}
%\begin{tabular}{c c}
%\textbf{(a)} & \textbf{(b)} \\
       % \includegraphics[scale=0.8]{EOS1.pdf} & \includegraphics[scale=0.8]{EOSsloth1.pdf}
   % \end{tabular}
\caption{(a) (left panel) Variations of dark energy equation of state parameter with redshift for general quintessence dark energy model, considering different initial values of $\frac{\phi}{f}$. (b) (right panel) Variations of dark energy equation of state parameter with redshift for Slotheon field dark energy model, considering different initial values of $\frac{\pi}{f}$.} \label{eos2}
\end{center}
\end{figure}

In Fig. \ref{eos2} we plot the evolutions of dark energy equation of state $\omega_{\rm DE}$ as a function of redshift $z$ for different initial values of the field ($\phi_i$ or $\pi_i$). We consider in Fig. \ref{eos2}(a) the quintessence dark energy model and in Fig. \ref{eos2}(b) the Slotheon field dark energy model. For both the cases it can be noted that for smaller values of $\frac{\phi_i}{f}$ (or $\frac{\pi_i}{f}$) the evolutions of $\omega_{\rm DE}$ appears to resemble more to that for $\Lambda$CDM. Thus when initially the field is nearer to the top of the potential (from $V_{\rm DE}(\phi)$ and $V_{\rm DE}(\pi)$ it can be clearly observed that the potentials have their maximum value for $\frac{\phi_i}{f}$=0 or $\frac{\pi_i}{f}=0$ respectively), it will feel the steepness of the potential less severely and heads towards a slower rolling. We observe that if $\phi_i \simeq0$ (or  $\pi_i \simeq0$) then $\omega_{\rm DE}$ is almost equal to $-1$ all through, i.e., the field may not experience the slope of the potential and the effective dynamics is independent of $f$.

\begin{figure}[H]
\begin{center}
\includegraphics[scale=0.77]{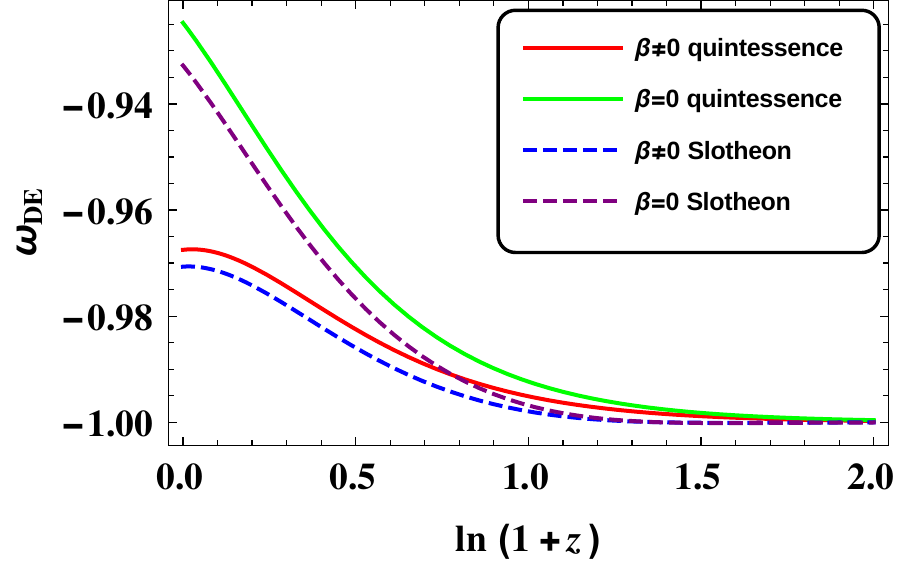}
\caption{Variations of dark energy equation of state parameters with redshift for quintessence dark energy (solid lines) and Slotheon field dark energy (dashed lines). Comparison between these two models of dark energy are shown in the presence of DM-DE interactions as also in the absence of DM-DE interactions.} \label{eos3}
\end{center}
\end{figure}

In Fig. \ref{eos3}, we plot the variations of the dark energy equation of state parameter $\omega_{\rm DE}$ with redshift for both the dark energy models namely quintessence (the solid lines) and the Slotheon (dashed lines). It can be easily observed from the graph that the behaviour of $\omega_{\rm DE}$ for the  Slotheon model behaviour is closer to the same for $\Lambda$CDM model than what is obtained for the quintessence model. This is expected as the Slotheon field favours the slow roll criteria more than the quintessence field, which we have discussed earlier. It is also interesting to note here that for both the cases when interactions between dark matter and dark energy are considered ($\beta \neq 0$), the behaviours of the fields resemble more to the behaviour of $\Lambda$CDM model. In other words the fields better satisfy the slow rolling criteria.

\begin{figure}[H]
%\begin{center}
\includegraphics[scale=0.8]{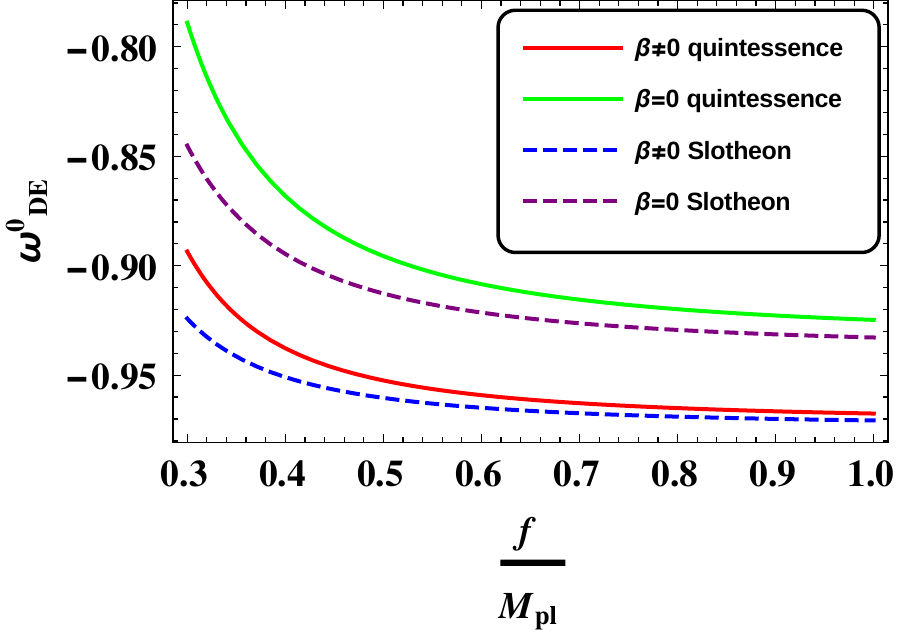}
\includegraphics[scale=0.8]{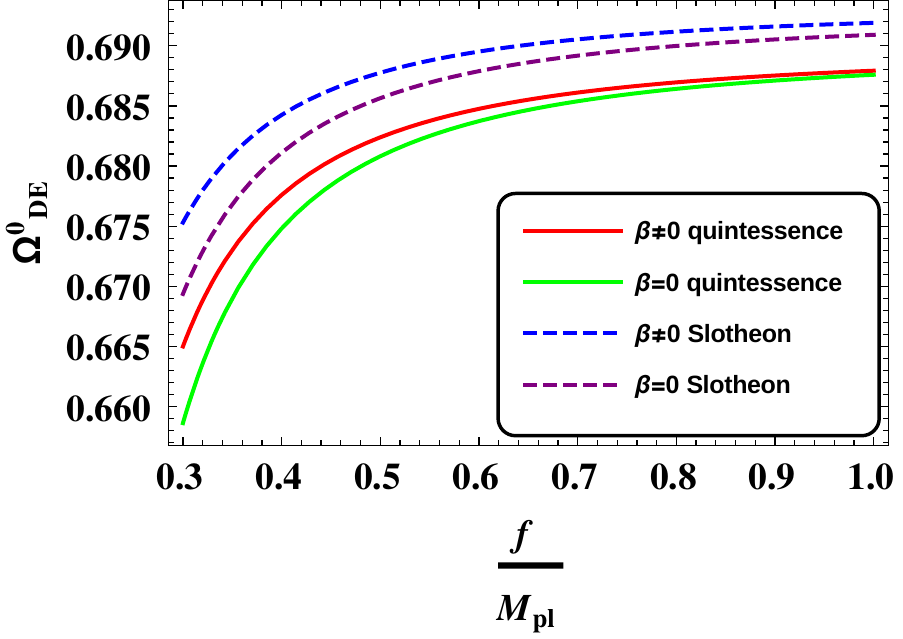}
%\begin{tabular}{c c}
%\textbf{(a)} & \textbf{(b)} \\
 %       \includegraphics[scale=0.8]{EOSkappa1.pdf} & \includegraphics[scale=0.8]
  %  \end{tabular}
\caption{(a) (left panel) Variations of the present value (at $z=0$ or $a=1$) of the dark energy equation of state parameters $\omega_{\rm DE}^0$ with $\frac{f}{M_{\rm pl}}$ for quintessence dark energy model (solid lines) and Slotheon field dark energy model (dashed lines). Effects of absence and presence of the DM-DE interaction are also shown for both the fields. (b) (right panel) Same as Fig. \ref{eos4}(a) but for present value of dark energy density parameter $\Omega_{\rm DE}^0$.  See text for detail discussion.}\label{eos4}
%\end{center}
\end{figure}

\begin{figure}[H]
%\begin{center}
\includegraphics[scale=0.8]{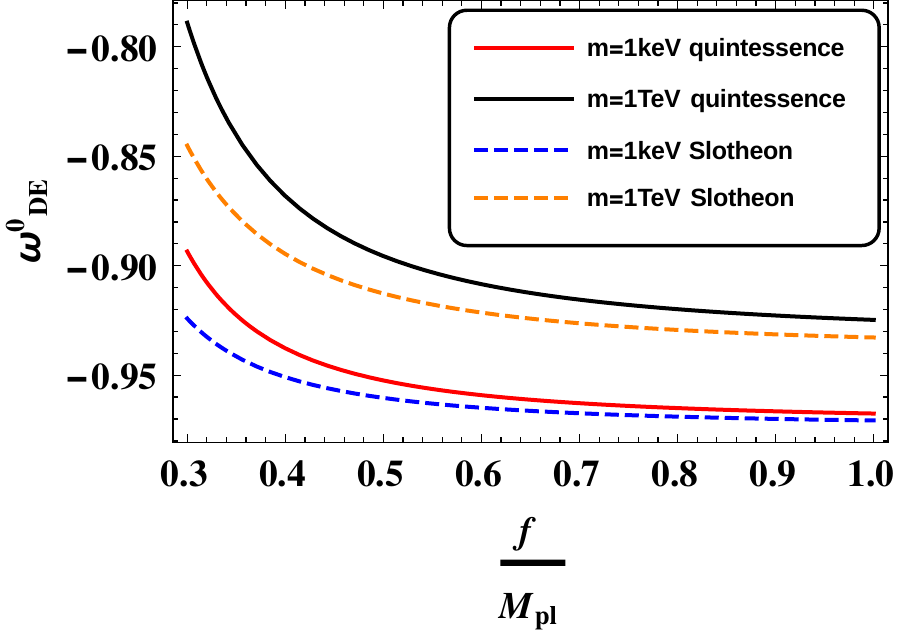}
\includegraphics[scale=0.8]{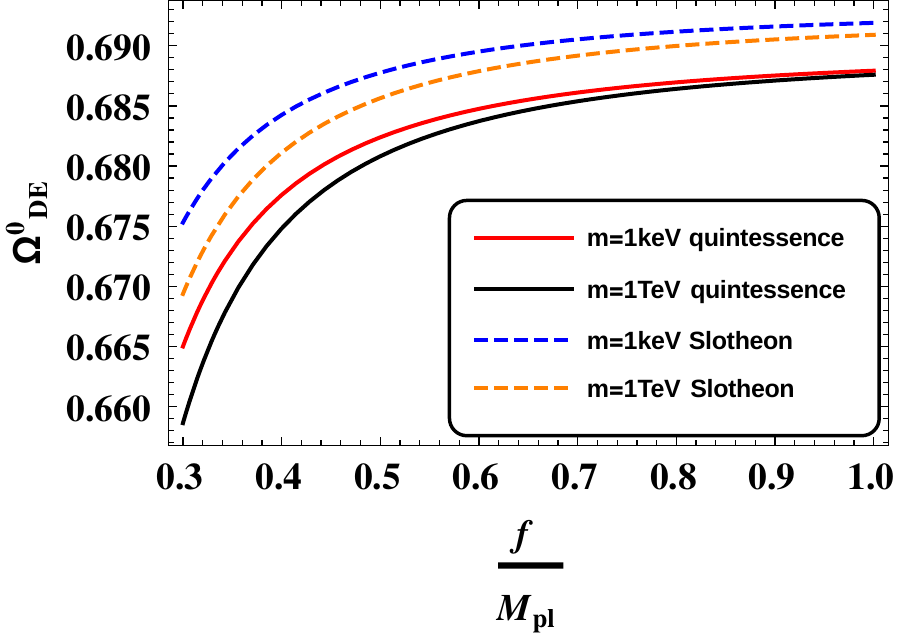}
\caption{(a) (left panel) Comparisons of evolutions of the present epoch value of dark energy equation of state parameter with $\frac{f}{M_{\rm pl}}$ for quintessence dark energy model (solid lines) and Slotheon field dark enegy model (dashed lines) for two chosen dark matter ``bare" masses namely $m=1$ keV and $m=1$ TeV. (b) (right panel) Same as Fig. \ref{eos5}(a) but for present value of dark energy density parameter $\Omega_{\rm DE}^0$.  See text for discussion.}\label{eos5}
%\end{center}
\end{figure}

In order to understand how sensitive the dark energy equation of state $\omega_{\rm DE}$ and the dark energy density parameter $\Omega_{\rm DE}$ at the present epoch are, to the variation of $f$, we calculate these variations using the formalism given in section 3 and section 4 and the results are plotted in Fig. \ref{eos4} (a and b) and in Fig. \ref{eos5} (a and b). 

In Fig. \ref{eos4}(a) (left panel of Fig. \ref{eos4}) the variations of the present value of the equation of state parameters $\omega_{\rm DE}^0$ (value of $\omega_{\rm DE}$ at redshift $z=0$) with the spontaneous symmetry breaking scale $f$ for both the models quintessence (marked with solid lines) and the Slotheon model (marked with dashed lines) are shown. We not only compare the variations for both the dark energy models considered here but also the results we obtain when DM-DE interaction ($\beta \neq 0$) is included in the calculations and the case when such interaction is not considered ($\beta=0$). We adopt the range of the present value of $\omega_{\rm DE}$ to be $-1.038\leqslant\omega_{\rm DE}^0\leqslant-0.884$ as given by PLANCK 2018 \cite{planck}. It is to be noted that in all the four cases shown in Fig. \ref{eos4}(a) the calculated values of $\omega_{\rm DE}^0$ lie well within the range of PLANCK for higher values of $f$ while the former deviates from this range for smaller values of the scale $f$. This is due to the fact that since $f$ is associated with a steepness of the pNGB potentials, pNGB potential tends to be flat as $f$ increases. A discussion is in order. For the quintessence case with $\beta=0$ the value of $\omega_{\rm DE}^0$ goes beyond the PLANCK range at $f < 0.4 M_{\rm pl}$ while for $f\geqslant0.4 M_{\rm pl}$, $\omega_{\rm DE}^0$ remains barely within the PLANCK range. The situation is much improved when DM-DE interaction is switched on ($\beta \neq 0$). In this situation $\omega_{\rm DE}^0$ lies well within the PLANCK range even upto $f \sim 0.3 M_{\rm pl}$. A similar situation is observed for the Slotheon case too. But as seen from Fig. \ref{eos4}(a), Slotheon results are always better than those obtained from quintessence since for both $\beta=0$ and $\beta\neq 0$, the range of $f$ for which the $\omega_{\rm DE}^0$ agrees with the PLANCK limit are always larger for the latter case. Similar conclusions can be drawn from Fig. \ref{eos4}(b) (right panel of Fig. \ref{eos4}) too where present value of dark energy density parameter $\Omega_{\rm DE}^0$ (value of $\Omega_{\rm DE}$ at $z=0$) is plotted for various $f$ values and compare with PLANCK range given by \cite{planck} $0.678\leqslant\Omega_{\rm DE}^0\leqslant0.692$. In addition one also note that in case of $\Omega_{\rm DE}^0$ (Fig. \ref{eos4}(b)) the Slotheon field results for both $\beta=0$ and $\beta \neq 0$ are in better agreement with the PLANCK limit than those for quintessence considerations. Therefore from Fig. \ref{eos4}(a) and Fig. \ref{eos4}(b) we may conclude that the Slotheon field dark energy model with DM-DE interactions address the higher-$f$ problem most effectively. 

In order to study how the nature of variation of $\omega_{\rm DE}^0$ with $f$ changes for different ``bare" dark matter masses $m$ in presence of DM-DE interaction ($\beta \neq 0$) we repeat the analyses shown in Fig. \ref{eos4} for different choices of $m$. We found that when $m < 1$ keV the values of $\omega_{\rm DE}^0$ for the chosen range of $\frac{f}{M_{\rm pl}}$ do not lie within the PLANCK limit for $\omega_{\rm DE}^0$. Again for $m > 1$ TeV all $\omega_{\rm DE}^0$ values for the same choice of $\frac{f}{M_{\rm pl}}$ range lie beyond the PLANCK limit for $\omega_{\rm DE}^0$. In Fig. \ref{eos5} (a and b) we plot $\omega_{\rm DE}^0$ vs $\frac{f}{M_{\rm pl}}$ (Fig. \ref{eos5}(a)) and $\Omega_{\rm DE}^0$ vs $\frac{f}{M_{\rm pl}}$ (Fig. \ref{eos5}(b)) for the cases of Slotheon field and quintessence field for two values of ``bare" dark matter masses namely 1 TeV and 1 keV and compare the results. 

It is obvious that similar trend as in Fig. \ref{eos4} is reflected in Fig. \ref{eos5} too. Although even for small $f$-values the PLANCK result is satisfied for $\omega_{\rm DE}^0$ for both the quintessence and Slotheon dark energy case when $m=1$ keV (Fig. \ref{eos5}(a)) but for Slotheon case the variation of $\omega_{\rm DE}^0$ lies below the quintessence case indicating that the Slotheon case satisfies the PLANCK limit better even if $f$ values are further lowered. For $m=1$ TeV (higher value) the PLANCK range is generally satisfied for both the Slotheon and quintessence cases when $f$ is high but in this case also more entered lower range of $f$ can be explored for Slotheon model results than those for quintessence. From Fig. \ref{eos5}(b), similar conclusions can be drawn by observing how the variations of $\Omega_{\rm DE}^0$ with $f$ obey the PLANCK range. Therefore from Fig. \ref{eos5}(a) and Fig. \ref{eos5}(b) it is observed that the Slotheon field dark energy when coupled to dark matter of ``bare" mass $\sim 1$ keV, can approach the high-$f$ problem most effectively.

\section{Summary and Discussions}
In this work we explore the dark energy equation of state and dark energy density parameter in case of a dark matter - dark energy interaction where the dark energy is considered to have driven by a pNGB scalar field $\alpha$ with potential having a form $V(\alpha)=\mu^4(1+{\rm cos}(\frac{\alpha}{f}))$. A pNGB is produced when the Nambu Goldstone boson that arises due to spontaneous breaking of a global symmetry acquires a mass at soft explicit symmetry breaking scale lower than the spontaneous symmetry breaking scale $f$. We address in this work the high-$f$ problem of such dark energy models. The high-$f$ problem arises from the consideration that although large value of $f$ leads to a flatter potential, but it is not compatible with the field theory limits and quantum gravity corrections. In this work we show that when dark matter - dark energy interaction is taken into account, the calculated cosmological parameters agree much better with the observational results of PLANCK 2018 experiment for lower values of $f$ and then the high-$f$ problem  can be averted. We show this for both the quintessence dark energy model and for another model namely the Slotheon dark energy model. The latter is inspired by the theories of extra dimensions such as the DGP theory. We also find that Slotheon dark energy model with the DM-DE interaction addresses better the high-$f$ problem than the quintessence model. 

We also explore in the present framework of dark matter - dark energy interaction, the mass limits of dark matter that would produce dark energy equation of state and dark energy density parameters for low values of $f$. We found this limit to lie between $\sim 1$ keV and $\sim 1$ TeV. The PLANCK limit is found to be better satisfied for lower values of the mass of the dark matter (the lower limit being $\sim 1$ keV). Here too the Slotheon dark energy consideration appears to be better than the quintessence dark energy.
 
Differences among the three dark energy models we mention in this work (coupled quintessence dark energy model, coupled Slotheon dark energy model and $\Lambda$CDM model) can also be demonstrated by considering the statefinder parameters \cite{sahni}. The statefinder parameters $\{r,s\}$ serve as a geometrical diagnostic pair to probe dark energy and defined as \cite{sahni}
\begin{eqnarray}
r &=& \frac{\dddot{a}}{a H^3}\,\,,\\
s &=& \frac{r-1}{3(q-1/2)}\,\,,
\end{eqnarray} 
where decelaration parameter $q=-\frac{\ddot{a}}{a H^2}$. In the late time ($z<10^4$) when the Universe is dominated by its two components namely the dark energy component and the matter component, the pair $\{r,s\}$ can be expressed as \cite{dds}
\begin{eqnarray}
r &=& 1+\frac{9}{2}\Omega_{\rm DE}\omega_{\rm DE}(1+\omega_{\rm DE})-\frac{3}{2}a \Omega_{\rm DE}\frac{{d\omega}_{\rm DE}}{da}\,\,,\\
s &=& 1+\omega_{\rm DE} -\frac{1}{3}\frac{a}{\omega_{\rm DE}}\frac{d\omega_{\rm DE}}{da}\,\,.
\end{eqnarray}
For $\Lambda$CDM model, as $\omega_{\rm DE}=-1$ is constant throughout the evolution of the Universe the pair $\{r,s\}$ takes the value $r=1$ and $s=0$. Hence any deviation of $r$ from $1$ and $s$ from $0$ would show the deviation of the nature of dark energy from cosmological constant and would reveal the dynamical nature of dark energy. In Ref. \cite{ujjaini} authors showed that the quintessence models of dark energy are characterized by the values in the region $r < 1$ and $s > 0$.\\
In Ref. \cite{sahniom} Sahni et al proposed an alternative route to distinguished $\Lambda$CDM model from other dynamical dark energy models. They introduced a new parameter $Om(z)$ to distinguished different dark energy models and defined it as
\begin{equation}
Om(z)=\frac{h^2(z)-1}{(1+z)^3-1}\,\,,
\end{equation}
where $h(z)=H(z)/H_0$. For $\Lambda$CDM model $Om(z)=\Omega_{0m}$ (matter density parameter at $z=0$) while for quintessence dark energy ($\omega > -1$) $Om(z)>\Omega_{0m}$ whereas $Om(z)<\Omega_{0m}$ for phantom dark enrgy model ($\omega < -1$) \cite{sahniom}. Moreover in Ref. \cite{sahniom2} the authors described the two point $Om$ diagnostic and three pont $Om$ diagnostic in detail and show that different $Om$ diagnostics are suitable for analysing different cosmological observations.

\begin{figure}
\includegraphics[scale=0.8]{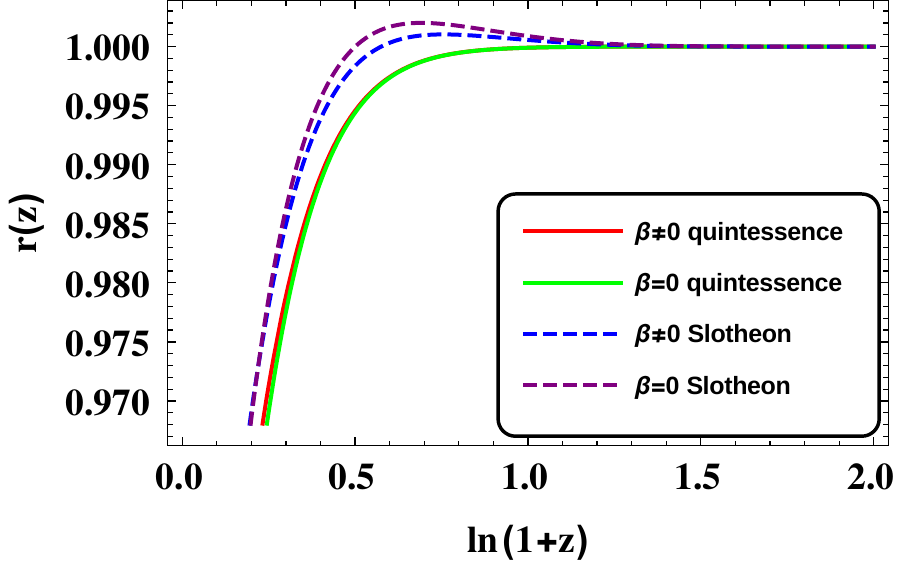} 
\includegraphics[scale=0.8]{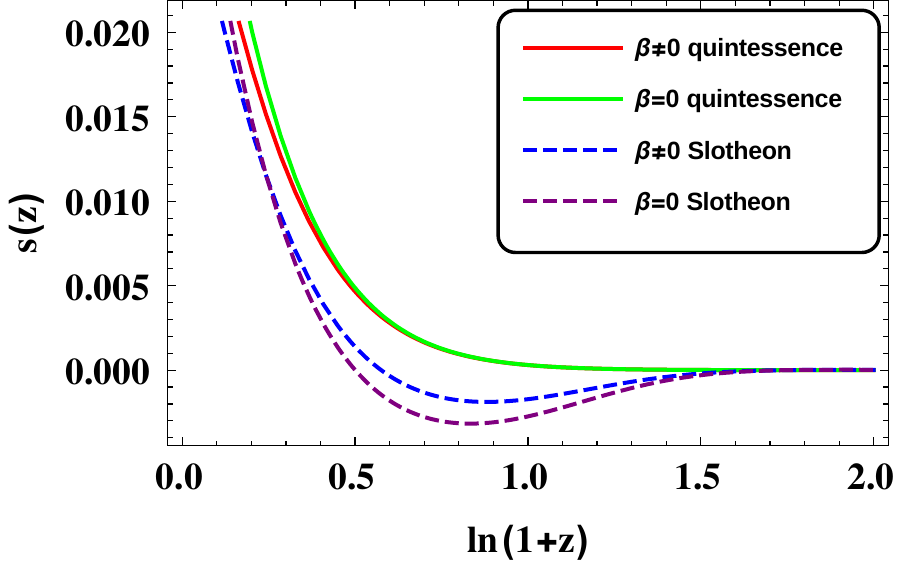}
\begin{center}
\includegraphics[scale=0.8]{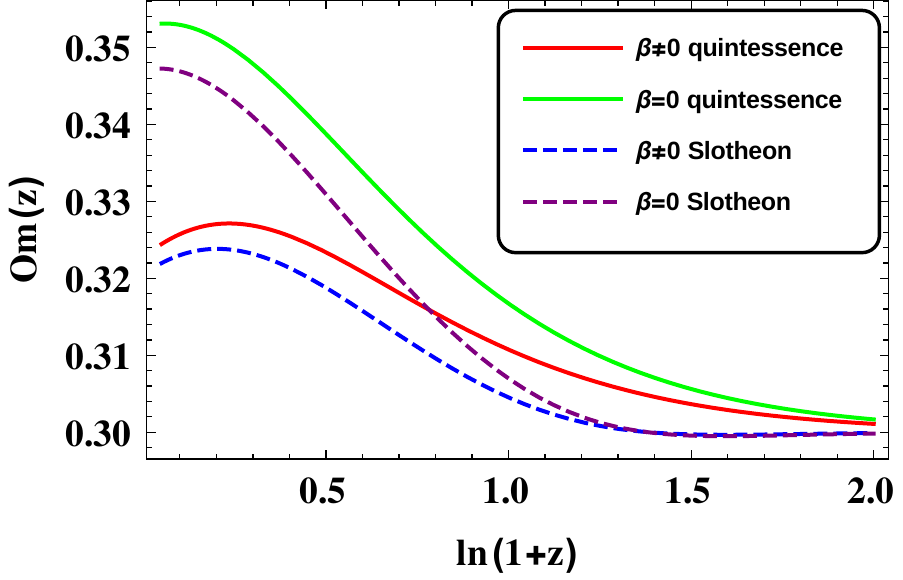}
\end{center}
\caption{Variations of $r$ with $z$ (top left), $s$ with $z$ (top right) and $Om(z)$ with $z$ (bottom) for four different cases, namely minimally coupled quintessence model ($\beta=0$, solid green line), coupled quintessence model ($\beta \neq 0$, solid red line), minimally coupled Slotheon dark energy model ($\beta=0$, dashed purple line) and coupled Slotheon dark energy model ($\beta\neq 0$, dashed blue line).}\label{rsom}
\end{figure}

In Fig. \ref{rsom} we plot the variations of the statefinder parameter $r$ and $s$ with redshift $z$ and the variations of $Om(z)$ with $z$ for four different cases, namely minimally coupled quintessence model ($\beta=0$, solid green line), coupled quintessence model ($\beta \neq 0$, solid red line), minimally coupled Slotheon dark energy model ($\beta=0$, dashed purple line) and coupled Slotheon dark energy model ($\beta\neq 0$, dashed blue line). From the above discussion we have for $\Lambda$CDM $r=1$, $s=0$ and $Om(z)=\Omega_{0m} \simeq 0.3$. Therefore the plots in Fig. \ref{rsom} clearly reveal the dynamical nature of these dark energy models. The plots (Fig. \ref{rsom}) also indicate the thawing nature \cite{thawing} of dark energy models i.e., the models are identical with $\Lambda$CDM at early time but gradually deviate from it with time. The characteristics of quintessence dark energy models also are evident in Fig. \ref{rsom} since we obtain $r<1$, $s>0$ and $Om(z) > \Omega_{0m}$ for quintessence  model. In Fig. \ref{rsom} the quintessence dark energy models and Slotheon dark energy models can also be distinguished from each other. Moreover it is clear from the plots in Fig. \ref{rsom} that statefinder pair $\{r,s\}$ and $Om(z)$ diagnostic can distinguish between minimally coupled dark energy models ($\beta=0$) and non-minimally coupled dark energy models ($\beta \neq 0$).

It may also be mentioned that the Slotheon dark energy model can not be ruled out from the LIGO detections of Gravitational Waves. From the LIGO observation of Gravitational Wave (GW) arising out of the neutron star merger (GW 170817), the GW speed is constrained as \cite{GW}
$$
-3 \times 10^{-15} \leq (c_T/c -1) \leq 7 \times 10^{-16},
$$
where $c_T$ denotes the GW phase velocity and $c$ the speed of light which implies luminal speed of  GW. The operational frequency of LIGO for the above detection of GW is 10-100 Hz \cite{GW}. In the same reference \cite{GW} this is also summarised that generically speaking the GW speed $c_T$ is obtained at the frequency it is measured. Therefore  the GW speed at the measured frequency could be the constraint on the GW speed obtained from an effective field theory (EFT) such as Horndeski theory \cite{horndeski} of dark energy at that frequency. The Slotheon model of dark energy which arises out of the decoupling limit of DGP model also belongs to the Horndeski EFT model. Such a DGP theory can have UV completion \cite{UV} and the sound speed remains luminal without being influenced by the background configuration. The theory can yield luminal GW speed at LIGO at the frequency of LIGO \cite{GW}. Thus Slotheon dark energy model also could produce luminal GW speed at LIGO frequency and therefore could not be ruled out at present.

From the present analyses and calculations, this may be concluded that the high-$f$ value problem for pNGB potential can be better addressed if dark matter - dark energy interaction is present. Also the Slotheon dark energy model is better suited than the quintessence model for the purpose. In addition dark matter mass limit for such a scenario is estimated.
 \vskip 1cm
\noindent{\large \bf Acknowledgements}  

\noindent The authors would like to thank A. Biswas and A. Dutta Banik for some useful suggestions. One of the authors (U.M.) receives her fellowship (as a graduate student leading to Ph.D. degree) grant from Council of Scientific \& Industrial Research (CSIR), Government of India as Junior Research Fellow (JRF) having the fellowship Grant No. 09/489(0106)/2017-EMR-I.
%\begin{center}


\begin{thebibliography}{99}
%\cite{observation1}
\bibitem{observation1}
A.G. Riess et al., Supernova Search Team collaboration, Astron. J. {\bf 116}, 1009 (1998).

\bibitem{Spergel:2003}
 D.N. Spergel et al., WMAP collaboration, Astrophys. J. Suppl. {\bf 148}, 175 (2003).

\bibitem{Delubac:2015}
 T. Delubac et al., BOSS collaboration, Astron. Astrophys. {\bf 574}, A59 (2015).

%\cite{observation2}
\bibitem{observation2}
E. Hawkins et al., Mon. Not. Roy. Astron. Soc. {\bf 346}, 78 (2003).

\bibitem{lambda CDM}
E.J. Copeland, M. Sami and S. Tsujikawa, Int. J. Mod. Phys. D {\bf 15}, 1753 (2006).

\bibitem{kazuharu_review}
 K.~Bamba, S.~Capozziello, S.~Nojiri and S.~D.~Odintsov,
  %``Dark energy cosmology: the equivalent description via different theoretical models and cosmography tests,''
  Astrophys.\ Space Sci.\  {\bf 342}, 155 (2012).

%\bibitem{fine tuning}

\bibitem{cosmic coincidence}
 P.J. Steinhardt, L.-M. Wang and I. Zlatev, Phys. Rev. D {\bf 59}, 123504 (1999).

\bibitem{quint1}
 S.  Tsujikawa,  Class.  Quant.  Grav. {\bf 30},  214003  (2013).
 %,arXiv:1304.1961 [gr-qc].

\bibitem{quint2}
R. R. Caldwell and R. Dave and P. J. Steinhardt, Phys.Rev.  Lett. {\bf 80},  1582 (1998).
%,  arXiv:astro-ph/9708069[astro-ph].


\bibitem{extra dim1}
L. Randalland R. Sundrum, Phys. Rev. Lett. {\bf 83} , 4690 (1999).


\bibitem{extra dim2}
M.R. Garousi, M. Sami and S. Tsujikawa, Phys. Rev.D {\bf 70}, 04353 (2004).

%\cite{dm1}
\bibitem{dm1} 
  D.~E.~McLaughlin,
  %``Evidence in virgo for the universal dark matter halo,''
  Astrophys.\ J.\  {\bf 512}, L9 (1999).
%  doi:10.1086/311860
%  [astro-ph/9812242].
  %%CITATION = doi:10.1086/311860;%%
  %85 citations counted in INSPIRE as of 27 Aug 2019
 
 \bibitem{Lokas}  
E L Lokas and G A Mamon, Mon. Not. R. Astron. Soc. {\bf 343}, 401 (2003).
%, arXiv:astro-ph/0302461


\bibitem{dm2} 
M Bradac, Nucl. Phys. Proc. Suppl. {\bf 194}, 17 (2009).


%\cite{Boehm:2003hm}


\bibitem{Wang:2016}
B. Wang, E. Abdalla, F. Atrio-Barandela and D. Pavon
%, Dark Matter and Dark Energy Interactions: Theoretical Challenges, Cosmological Implications and Observational Signatures,
Rept. Prog. Phys. {\bf 79}, 096901 (2016).
%096901, [1603.08299].


\bibitem{Farrar:2004} 
 G. R. Farrar and P. J. E. Peebles,
%  Interacting dark matter and dark energy,
Astrophys. J. {\bf 604}, 1 (2004).
%1–11, [astro-ph/0307316].
 
 \bibitem{Amendola:2007}
L. Amendola, G. Camargo Campos and R. Rosenfeld,
%Consequences of dark matter-dark energy interaction
%on cosmological parameters derived from SNIa data,
Phys. Rev. D {\bf 75}, 083506 (2007).
% [astro-ph/0610806].
 
\bibitem{He:2008}
J.-H. He and B. Wang,
% Effects of the interaction between dark energy and dark matter on cosmological parameters,
JCAP {\bf 06}, 010 (2008).
%, [0801.4233].

\bibitem{Caldera:2009}
G. Caldera-Cabral, R. Maartens and B. M. Schaefer,
%The Growth of Structure in Interacting Dark Energy Models,
JCAP {\bf 07}, 027 (2009).
 %, [0905.0492].

\bibitem{Abdalla:2010}
E. Abdalla, L. R. Abramo and J. C. C. de Souza,
%Signature of the interaction between dark energy and dark matter in observations, 
Phys. Rev. D {\bf 82},  023508 (2010).
%, [0910.5236].

\bibitem{Pan:2015}
S. Pan, S. Bhattacharya and S. Chakraborty, 
%An analytic model for interacting dark energy and its observational constraints,
Mon. Not. Roy. Astron. Soc. {\bf 452}, 3038 (2015).
%3038–3046 , [1210.0396].

\bibitem{Tamanini:2015}
N. Tamanini,
%Phenomenological models of dark energy interacting with dark matter,
Phys. Rev. D {\bf 92}, 043524 (2015).
%043524, [1504.07397].

\bibitem{Sharov:2017}
G. S. Sharov, S. Bhattacharya, S. Pan, R. C. Nunes and S. Chakraborty,
%A new interacting two fluid model and its consequences,
Mon. Not. Roy. Astron. Soc. {\bf 466}, 3497 (2017).
% 3497–3506, [1701.00780].

\bibitem{Kumar:2017}
S. Kumar and R. C. Nunes,
% Observational constraints on dark matter–dark energy scattering cross section,
Eur. Phys. J. C {\bf 77}, 734 (2017).
% 734, [1709.02384].

\bibitem{Costa:2018}
A. A. Costa, R. C. G. Landim, B. Wang and E. Abdalla,
% Interacting Dark Energy: Possible Explanation for 21-cm Absorption at Cosmic Dawn,
Eur. Phys. J. C {\bf 78}. 746 (2018).
% 746, [1803.06944].



\bibitem{Yang:2019}
W. Yang, S. Vagnozzi, E. Di Valentino, R. C. Nunes, S. Pan and D. F. Mota,
% Listening to the sound of dark sector interactions with gravitational wave standard sirens, 
JCAP {\bf 07}, 037 (2019).
% 037, [1905.08286].

\bibitem{} A.~B.~Balakin and A.~S.~Ilin,
%``Dark Energy and Dark Matter Interaction: Kernels of Volterra Type and Coincidence Problem,''
Symmetry \textbf{10}, 411 (2018).

\bibitem{} M.~Khurshudyan and A.~Khurshudyan,
%``Some interacting dark energy models,''
Symmetry \textbf{10}, no.11, 577 (2018).

\bibitem{} S. Chakrabarti and S. Chattopadhyay, Zeitschrift fÃ¼r
Naturforschung A, \textbf{73}, 251-257 (2018).

%\cite{DiValentino:2019ffd}
\bibitem{DiValentino:2019ffd} 
  E.~Di Valentino, A.~Melchiorri, O.~Mena and S.~Vagnozzi,
  %``Interacting dark energy after the latest Planck, DES, and $H_0$ measurements: an excellent solution to the $H_0$ and cosmic shear tensions,''
  arXiv:1908.04281 [astro-ph.CO].
  %%CITATION = ARXIV:1908.04281;%%
  %1 citations counted in INSPIRE as of 27 Aug 2019.

\bibitem{267_sami}
C. F. Kolda and D. H. Lyth, Phys. Lett. B {\bf 458}, 197(1999).


\bibitem{279_sami}
J. A. Frieman, C. T. Hill, A. Stebbins and I. Waga,Phys. Rev. Lett. {\bf 75}, 2077 (1995).

\bibitem{280_sami}
K. Choi, Phys. Rev. D {\bf 62}, 043509 (2000).

%\bibitem{axion}
%\cite{Weinberg:1977ma}
%\bibitem{Weinberg:1977ma} 
 % S.~Weinberg,
  %``A New Light Boson?,''
  %Phys.\ Rev.\ Lett.\  {\bf 40}, 223 (1978).
%  doi:10.1103/PhysRevLett.40.223
  %%CITATION = doi:10.1103/PhysRevLett.40.223;%%
  %2916 citations counted in INSPIRE as of 10 Jan 2018

%\cite{Peccei:2006as}
\bibitem{Peccei:2006as} 
  R.~D.~Peccei,
  %``The Strong CP problem and axions,''
  Lect.\ Notes Phys.\  {\bf 741}, 3 (2008).
%  doi:10.1007/978-3-540-73518-2\_1
%  [hep-ph/0607268].
  %%CITATION = doi:10.1007/978-3-540-73518-2_1;%%
  %218 citations counted in INSPIRE as of 10 Jan 2018

%\cite{Peccei:1977hh}
\bibitem{Peccei:1977hh} 
  R.~D.~Peccei and H.~R.~Quinn,
  %``CP Conservation in the Presence of Instantons,''
  Phys.\ Rev.\ Lett.\  {\bf 38}, 1440 (1977).
%  doi:10.1103/PhysRevLett.38.1440
  %%CITATION = doi:10.1103/PhysRevLett.38.1440;%%
  %4193 citations counted in INSPIRE as of 10 Jan 2018

\bibitem{khlopov}
M.~Y.~Khlopov, A.~S.~Sakharov and D.~D.~Sokoloff,
  %``The nonlinear modulation of the density distribution in standard axionic CDM and its cosmological impact,''
  Nucl.\ Phys.\ Proc.\ Suppl.\  {\bf 72}, 105 (1999).


\bibitem{carrol}
S.M. Carroll, Phys. Rev. Lett. {\bf 81}, 3067 (1998).

\bibitem{Frieman}
Frieman J A, Hill C T, Stebbins A and Waga I, Phys. Rev. Lett. {\bf 75}, 2077 (1995).

%\bibitem{Nomura}
%Nomura Y, Watari T and Yanagida T, Phys. Lett. B {\bf 484}, 103 (2000).


\bibitem{Kim}
Kim J E and Nilles H P, Phys. Lett. B {\bf 553}, 1 (2003).


%\cite{ArkaniHamed:2003mz}
\bibitem{ArkaniHamed:2003mz} 
  N.~Arkani-Hamed, H.~C.~Cheng, P.~Creminelli and L.~Randall,
  %``Pseudonatural inflation,''
  JCAP {\bf 07}, 003 (2003).
%  doi:10.1088/1475-7516/2003/07/003
%  [hep-th/0302034].
  %%CITATION = doi:10.1088/1475-7516/2003/07/003;%%
  %127 citations counted in INSPIRE as of 28 Aug 2019

%\cite{Kaloper:2005aj}
\bibitem{Kaloper:2005aj} 
  N.~Kaloper and L.~Sorbo,
  %``Of pngb quintessence,''
  JCAP {\bf 0604}, 007 (2006).
%  doi:10.1088/1475-7516/2006/04/007
 % [astro-ph/0511543].
  %%CITATION = doi:10.1088/1475-7516/2006/04/007;%%
  %44 citations counted in INSPIRE as of 28 Aug 2019

%\cite{Adak:2014moa}
\bibitem{Adak:2014moa} 
  D.~Adak and K.~Dutta,
  %``Viable dark energy models using pseudo-Nambu-Goldstone bosons,''
  Phys.\ Rev.\ D {\bf 90}, 043502 (2014).
 % doi:10.1103/PhysRevD.90.043502
%  [arXiv:1404.1570 [astro-ph.CO]].
  %%CITATION = doi:10.1103/PhysRevD.90.043502;%%
  %3 citations counted in INSPIRE as of 28 Aug 2019

\bibitem{me_sloth}
 U.~Mukhopadhyay, D.~Majumdar and D.~Adak,
  %``Evolution of Dark Energy Perturbations for Slotheon Field and Power Spectrum,''
  arXiv:1903.08650 [gr-qc].


\bibitem{deba_sloth}
D.~Adak, A.~Ali and D.~Majumdar,
  %``Late time acceleration in a slow moving galileon field,''
  Phys.\ Rev.\ D {\bf 88}, 024007 (2013).
  

\bibitem{178_thesis}
C. Germani, L. Martucci, and P. Moyassari,
% Introducing the Slotheon: a slow Galileon scalar field in curved space-time, 
Phys.Rev. D {\bf 85}, 103501, (2012).


\bibitem{DGP}
  G. Dvali, G. Gabadadze, and M. Porrati, Phys. Lett. B {\bf 485}, 208 (2000).

\bibitem{173_thesis}
M. A. Luty, M. Porrati, and R. Rattazzi, 
%Strong interactions and stability in the DGP model,
 JHEP {\bf 09}, 029 (2003).


\bibitem{176_thesis}B. Jain and J. Khoury,
% Cosmological Tests of Gravity,
Annals Phys. {\bf 325}, 1479, (2010).


\bibitem{me_swamp}
 U.~Mukhopadhyay and D.~Majumdar,
  %``Swampland criteria in the slotheon field dark energy,''
  Phys.\ Rev.\ D {\bf 100}, 024006 (2019).


\bibitem{gal_swamp}
 S.~Brahma and M.~W.~Hossain,
  %``Dark energy beyond quintessence: Constraints from the swampland,''
  JHEP {\bf 06}, 070 (2019).


\bibitem{gal_pert}
 B.R. Dinda, Md. W. Hossain and A.A. Sen, JCAP {\bf 01}, 045 (2018).


\bibitem{usual_Q1}
%\cite{Bertolami:2007zm}
%\bibitem{Bertolami:2007zm} 
  O.~Bertolami, F.~Gil Pedro and M.~Le Delliou,
  %``Dark Energy-Dark Matter Interaction and the Violation of the Equivalence Principle from the Abell Cluster A586,''
  Phys.\ Lett.\ B {\bf 654}, 165 (2007).
%  doi:10.1016/j.physletb.2007.08.046
%  [astro-ph/0703462 [ASTRO-PH]].
  %%CITATION = doi:10.1016/j.physletb.2007.08.046;%%
  %272 citations counted in INSPIRE as of 28 Aug 2019

%\cite{Guo:2007zk}
\bibitem{usual_Q2}
%\bibitem{Guo:2007zk} 
  Z.~K.~Guo, N.~Ohta and S.~Tsujikawa,
  %``Probing the Coupling between Dark Components of the Universe,''
  Phys.\ Rev.\ D {\bf 76}, 023508 (2007).
%  doi:10.1103/PhysRevD.76.023508
%  [astro-ph/0702015 [ASTRO-PH]].
  %%CITATION = doi:10.1103/PhysRevD.76.023508;%%
  %315 citations counted in INSPIRE as of 28 Aug 2019


\bibitem{interacting_quint1}
%\cite{Wetterich:1994bg}
%\bibitem{Wetterich:1994bg} 
  C.~Wetterich,
  %``The Cosmon model for an asymptotically vanishing time dependent cosmological 'constant',''
  Astron.\ Astrophys.\  {\bf 301}, 321 (1995).
 % [hep-th/9408025].
  %%CITATION = HEP-TH/9408025;%%
  %762 citations counted in INSPIRE as of 28 Aug 2019

\bibitem{interacting_quint2}
%\cite{Amendola:1999er}
%\bibitem{Amendola:1999er} 
  L.~Amendola,
  %``Coupled quintessence,''
  Phys.\ Rev.\ D {\bf 62}, 043511 (2000).
%  doi:10.1103/PhysRevD.62.043511
%  [astro-ph/9908023].
  %%CITATION = doi:10.1103/PhysRevD.62.043511;%%
  %1276 citations counted in INSPIRE as of 28 Aug 2019


\bibitem{44_lisboa}
%\cite{Zimdahl:2001ar}
%\bibitem{Zimdahl:2001ar} 
  W.~Zimdahl and D.~Pavon,
  %``Interacting quintessence,''
  Phys.\ Lett.\ B {\bf 521}, 133 (2001).
%  doi:10.1016/S0370-2693(01)01174-1
%  [astro-ph/0105479].
  %%CITATION = doi:10.1016/S0370-2693(01)01174-1;%%
  %482 citations counted in INSPIRE as of 28 Aug 2019
  
  
%\cite{Farrar:2003uw}
\bibitem{Farrar:2003uw} 
  G.~R.~Farrar and P.~J.~E.~Peebles,
  %``Interacting dark matter and dark energy,''
  Astrophys.\ J.\  {\bf 604}, 1 (2004).
 % doi:10.1086/381728
 % [astro-ph/0307316].
  %%CITATION = doi:10.1086/381728;%%
  %369 citations counted in INSPIRE as of 28 Aug 2019
 
   
 %\cite{Micheletti:2009pk}
\bibitem{Micheletti:2009pk} 
  S.~Micheletti, E.~Abdalla and B.~Wang,
  %``A Field Theory Model for Dark Matter and Dark Energy in Interaction,''
  Phys.\ Rev.\ D {\bf 79}, 123506 (2009).
%  doi:10.1103/PhysRevD.79.123506
%  [arXiv:0902.0318 [gr-qc]].
  %%CITATION = doi:10.1103/PhysRevD.79.123506;%%
  %76 citations counted in INSPIRE as of 28 Aug 2019

\bibitem{PRD_formalism}
 O.~Bertolami, P.~Carrilho and J.~Paramos,
  %``Two-scalar-field model for the interaction of dark energy and dark matter,''
  Phys.\ Rev.\ D {\bf 86}, 103522 (2012). 
 
%\bibitem{lisboa}
%\bibitem{oscillation}
%\cite{Turner:1983he}
%\bibitem{Turner:1983he} 
 % M.~S.~Turner,
  %``Coherent Scalar Field Oscillations in an Expanding Universe,''
  %Phys.\ Rev.\ D {\bf 28}, 1243 (1983).
%  doi:10.1103/PhysRevD.28.1243
  %%CITATION = doi:10.1103/PhysRevD.28.1243;%%
  %545 citations counted in INSPIRE as of 28 Aug 2019
  
 % \bibitem{lisboa}
%P. Carrilho, {\it A Scalar Field Theory for Dark Matter$-$Dark Energy Interaction}, Master's thesis, 2012, unpublished, https://fenix.tecnico.ulisboa.pt/downloadFile/
%395144546561/tese$\_$Pedro$\_$Carrilho$\_$63431.pdf.

\bibitem{24_swamp}
N. Chow and J. Khoury,Phys. Rev. D {\bf 80}, 024037 (2009).

\bibitem{25_swamp}
A. Ali, R.Gannouji, M. W. Hossain, and M. Sami,Phys.Lett. B {\bf 718}, 5 (2012).

\bibitem{177_thesis}
A. Nicolis, R. Rattazzi, and E. Trincherini,
% The Galileon as a local modification of gravity,
Phys.Rev. D {\bf 79},  064036 (2009).
%[arXiv:0811.2197].

%\cite{Amendola:1999er}
%\bibitem{Amendola:1999er} 
%  L.~Amendola,
  %``Coupled quintessence,''
 % Phys.\ Rev.\ D {\bf 62}, 043511 (2000).
  
  
  %\cite{Amendola:2000ub}
%\bibitem{Amendola:2000ub} 
 % L.~Amendola,
  %``Dark energy and the Boomerang data,''
  %Phys.\ Rev.\ Lett.\  {\bf 86}, 196 (2001).


%\bibitem{baryon_lucha}
%\cite{TocchiniValentini:2001ty}
%\bibitem{TocchiniValentini:2001ty} 
 % D.~Tocchini-Valentini and L.~Amendola,
  %``Stationary dark energy with a baryon dominated era: Solving the coincidence problem with a linear coupling,''
  %Phys.\ Rev.\ D {\bf 65}, 063508 (2002).
%  doi:10.1103/PhysRevD.65.063508
 % [astro-ph/0108143].
  %%CITATION = doi:10.1103/PhysRevD.65.063508;%%
  %134 citations counted in INSPIRE as of 28 Aug 2019


\bibitem{thawing}
R. R. Caldwell and E. V. Linder, Phys. Rev. Lett. {\bf 95}, 141301 (2005).



\bibitem{planck}
N. Aghanim et al., Planck Collaboration, arXiv:1807.06209 [astro-ph.CO].

%\bibitem{sfdmde}W.~Zimdahl and D.~Pavon,
%``Statefinder parameters for interacting dark energy,''
%Gen. Rel. Grav. \textbf{36}, 1483-1491 (2004).

\bibitem{sahni} V.~Sahni, T.~D.~Saini, A.~A.~Starobinsky and U.~Alam,
%``Statefinder: A New geometrical diagnostic of dark energy,''
JETP Lett. \textbf{77}, 201-206 (2003).

\bibitem{dds}D.~Adak, D.~Majumdar and S.~Pal,
%``Generalizing thawing dark energy models: the standard vis-à-vis model independent diagnostics,''
Mon. Not. Roy. Astron. Soc. \textbf{437}, no.1, 831-842.

\bibitem{ujjaini} U.~Alam, V.~Sahni, T.~D.~Saini and A.~A.~Starobinsky,
%``Exploring the expanding universe and dark energy using the Statefinder diagnostic,''
Mon. Not. Roy. Astron. Soc. \textbf{344}, 1057 (2003).

\bibitem{sahniom} V.~Sahni, A.~Shafieloo and A.~A.~Starobinsky,
%``Two new diagnostics of dark energy,''
Phys. Rev. D \textbf{78}, 103502 (2008).

\bibitem{sahniom2} A.~Shafieloo, V.~Sahni and A.~A.~Starobinsky,
%``A new null diagnostic customized for reconstructing the properties of dark energy from BAO data,''
Phys. Rev. D \textbf{86}, 103527 (2012).


\bibitem{GW} 
C. de Rham and S. Melville, Phys. Rev. Lett. {\bf 121}, 221101 (2018).

\bibitem{horndeski}
G.~W.~Horndeski,
  %``Second-order scalar-tensor field equations in a four-dimensional space,''
  Int.\ J.\ Theor.\ Phys.\  {\bf 10}, 363 (1974),
   N.~Franchini and T.~P.~Sotiriou,
  %``Cosmology with subdominant Horndeski scalar field,''
  arXiv:1903.05427 [gr-qc],
  L.~Heisenberg,
  %``A systematic approach to generalisations of General Relativity and their cosmological implications,''
  Phys.\ Rept.\  {\bf 796}, 1 (2019).

\bibitem{UV}
 R.~Gregory, N.~Kaloper, R.~C.~Myers and A.~Padilla,
  %``A New perspective on DGP gravity,''
  JHEP {\bf 0710}, 069 (2007).
\end{thebibliography}
\end{document}